\documentclass[a4paper,11pt]{article}
\usepackage{pos}
\usepackage{siunitx}
\usepackage{hyperref}
\usepackage[nameinlink]{cleveref}
\usepackage{wrapfig}

\def\xmax{$X_\mathrm{max}$ }
\def\nmax{$N_\mathrm{max}$ }
\def\lognmax{$\log _{10}(N_\mathrm{max})$ }
\def\vvb{$\vec{v} \times (\vec{v} \times \vec{B})$ }

\title{Beyond $X_\mathrm{max}$ : Reconstructing Air Shower Profiles with Information Field Theory with SKA-Low}
\ShortTitle{Air Shower Profile Reconstruction with IFT}

\author*[a]{K.~Watanabe}
\author[a,d]{T.~Huege}
\author[g,h,i]{T.A.~En{\ss}lin}
\author[g,h]{V.~Eberle}
\author[b]{S.~Bouma}
\author[c]{J.D.~Bray}
\author[d,e]{S.~Buitink}
\author[d,e]{A.~Corstanje} 
\author[d]{V.~De Henau}
\author[f]{E.~Dickinson}
\author[e]{T.~Gottmer}
\author[j,k]{B.~Hare}
\author[l]{H.~He}
\author[e]{J.R.~H\"orandel}
\author[f]{C.W.~James}
\author[g,h]{M.~Jetti}
\author[b]{P.~Laub}
\author[l]{X.~Li}
\author[j,k]{M.~Lourens}
\author[a]{H.J.~Mathes}
\author[e,m]{K.~Mulrey}
\author[b,n]{A.~Nelles}
\author[a,o]{S.~Saha}
\author[w]{F.~Schl{\"u}ter}
\author[j]{O.~Scholten}
\author[c]{R.E.~Spencer}
\author[k]{C.~Sterpka}
\author[k]{S.~ter Veen}
\author[b]{K.~Terveer}
\author[q]{T.N.G.~Trinh}
\author[j,k]{P.~Turekova}
\author[a]{D.~Veberi\v{c}}
\author[r]{M.~Waterson}
\author[s,t]{C.~Zhang}
\author[u]{P.~Zhang}
\author[l,v]{Y.~Zhang}

\emailAdd{keito.watanabe@kit.edu}

\abstract{

    While radio measurements of extensive air showers have shown to achieve a high precision of \xmax sensitivity, it has been shown that parameters beyond \xmax can also be reconstructed. These shape parameters contain additional sensitivity to the hadronic physics in the shower as well as its mass composition. In this work, we showcase a reconstruction framework to recover the full longitudinal profile from realistic radio measurements. The framework is based on Information Field Theory that infers the full profile with a forward-based model, which uses a Gaisser-Hillas profile with weakly informative shower priors, SMIET with a template library to synthesise pulses at any event geometry, and a realistic antenna response and noise level emulating that of SKA-Low. We verify the self-consistency of our framework with $\sim 900$ events generated with SMIET with antennas placed on the \vvb axis. The framework recovers the full profile within uncertainty and capture correlations between shower parameters. We yield an \xmax resolution of $< \SI{9}{\gram\per\centi\meter\squared}$ as well as resolutions of the width and asymmetry with minimal bias. The profile is also recovered with a bias of $< 4$\% at all atmospheric depths $< \SI{1200}{\gram\per\centi\meter\squared}$. We aim to apply this framework with pulses simulated from CoREAS with measured noise, ultimately extending the framework to realistic antenna layouts such as from LOFAR or SKA-Low. 
}

\FullConference{11th International Workshop on Acoustic and Radio EeV Neutrino Detection Activities (ARENA2026)\\
8-11 June 2026\\
Karlsruhe, Germany\\}

\begin{document}
\maketitle

\section{Introduction}

Radio measurements of extensive air showers produced from cosmic rays have shown significant development over the decades. In particular, several experiments have shown to achieve an \xmax sensitivity to high precision of $<\SI{20}{\gram\per\centi\meter\squared}$ \cite{Corstanje:2021kik, abdul_halim_radio_2024}, providing an alternative method for \xmax reconstruction alongside fluorescence- and particle-based detectors. In recent years, it has been shown that radio measurements are also sensitive to additional shower parameters that characterise the shape of the longitudinal evolution of the air shower (e.g. \cite{Buitink:2023reh, Corstanje:2023uyg}). These shape parameters have shown to contain sensitivity both to the mass composition of the cosmic ray, as well as the hadronic interactions within the shower. As such, reconstructing these additional parameters, along with $X_\mathrm{max}$, can aid us in determining the mass composition on an event level, as well as disentangling the dependence on hadronic interaction models used in Monte Carlo simulation tools for extensive air showers such as CORSIKA \cite{Heck:1998vt}. \par 

In our previous work \cite{watanabe_novel_2025}, we showcased a novel method to reconstruct the longitudinal evolution of the air shower directly from radio measurements using Information Field Theory \cite{Ensslin:2018pno} with SMIET \cite{Desmet:2025ufy} as part of our forward model. However, the model was specific to a particular event geometry and cosmic ray energy, and the prior distribution of shower parameters was directly dependent on the hadronic interaction model. In this work, we improve on the previous model with a more general forward-modelling of our framework, while still using realistic antenna models and noise characteristics based on the upcoming radio telescope SKA-Low, which will provide unprecedented accuracy in cosmic ray detection \cite{Huege:2026beb}. The self-consistency of the framework and intrinsic sensitivity of our reconstruction is verified using a SMIET-generated event dataset with varying event parameters and profiles.  

\section{Model}

The model used in this work is based on our previous work \cite{watanabe_novel_2025}, however with notable improvements. In this section, we describe the full model used for reconstructing the longitudinal profile, in particular highlighting changes made from our previous model. \par 

\textbf{Shower Profile: } The longitudinal profile of extensive air showers is well described by the Gaisser-Hillas function \cite{Gaisser:1977}. Here, we use the $LR$ formalism \cite{Andringa:2011zz}, motivated from previous studies that have already shown the mass- and hadronic model sensitivity of the shape parameters \cite{Buitink:2023reh, Corstanje:2023uyg}:
\begin{equation}
    N(X) = N_\mathrm{max} \exp\left(-\dfrac{X - X_\mathrm{max}}{RL}\right) \: \left(1 + \dfrac{R}{L} (X - X_\mathrm{max})\right)^{R^{-2}},
    \label{eq:gaisser_hillas}
\end{equation}
where $N_\mathrm{max}$ is the maximum number of electrons+positrons along the shower and \xmax is the atmospheric depth at $N_\mathrm{max}$. Here, $L$ and $R$ are the shape parameters that characterise the width and asymmetry of the shower profile, respectively. The profile is described as a function of the atmospheric depth, $X$, with units of $\si{\gram\per\centi\meter\squared}$. \par

\begin{wrapfigure}[9]{L}{0.42\textwidth}
    \sbox0{\small\begin{tabular}[t!]{lcc}
         \hline
         Parameter $p$ & $a_\mathrm{min}^p$ & $a_\mathrm{max}^p$ \\
         \hline
         \xmax & $\SI{400}{\gram\per\centi\meter\squared}$  & $\SI{1200}{\gram\per\centi\meter\squared}$ \\
         \lognmax & 7.0 & 9.0 \\
         $L$ & $\SI{180}{\gram\per\centi\meter\squared}$ & $\SI{300}{\gram\per\centi\meter\squared}$ \\
         $R$ & 0.2 & 0.5 \\
         \hline
    \end{tabular}}%
    \begin{minipage}[t!]{\wd0}
        \centering
        \usebox0
        \captionsetup{width=\wd0, justification=raggedright, singlelinecheck=false}
        \captionof{table}{Truncation limits used for $X_\mathrm{max}$, \lognmax, $L$, and $R$ for the truncated normal prior distribution.}
        \label{tab:shower_limits}
    \end{minipage}
\end{wrapfigure}

The shower parameters are known to have strong correlations between each other, in particular between $X_\mathrm{max}$, $L$, and $R$. In our previous model, we included the correlations explicitly as extracted from CORSIKA simulations from the LOFAR simulation library \cite{Corstanje:2021kik}. However, this limits and constrains the shape parameters to be sampled based on our informed correlations, which are also site- and atmosphere-dependent. Instead, we model all shower parameters using a truncated normal distribution, where the mean and standard deviation for each parameter $p$ are evaluated based on the truncation limits $[a_\mathrm{min}^p, a_\mathrm{max}^p]$: $\mu_p = (a_\mathrm{max}^p + a_\mathrm{min}^p) / 2$ and $\sigma_p = |a_\mathrm{max}^p - a_\mathrm{min}^p| / 4$, respectively. In this way, we allow the framework to implicitly pick up on the correlations instead. The limits on \xmax and \nmax are set by physically viable values, while the limits of $L$ and $R$ are chosen to exclude extreme values that are associated with anomalous showers. \Cref{tab:shower_limits} shows the limits chosen for all parameters. For \lognmax, we use a larger standard deviation of $\sigma_{\log_{10} N_\mathrm{max}} = 1.0$ instead, so as to not bias the amplitude strongly around the mean value. \par 

\begin{wrapfigure}[14]{R}{0.5\textwidth}
    \centering
    \includegraphics[width=\linewidth]{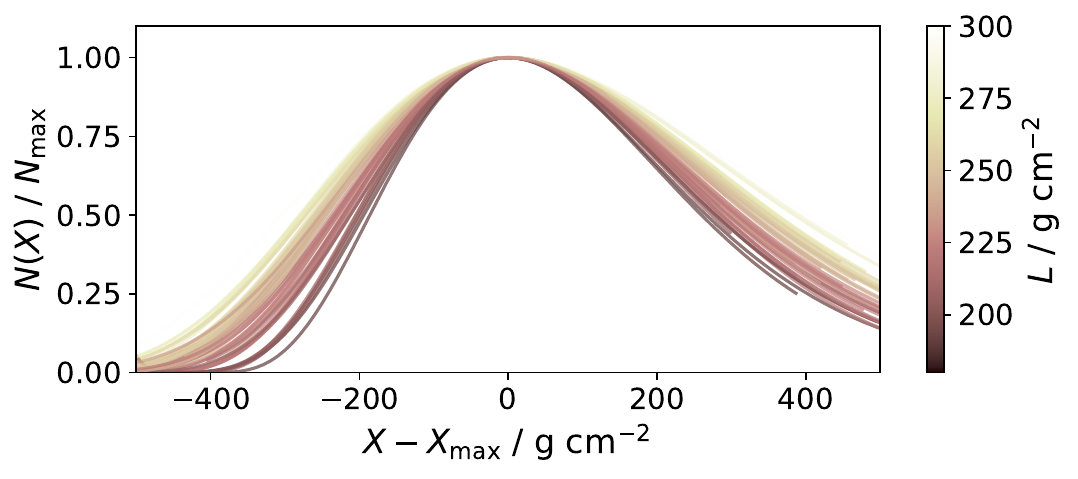}
    \caption{50 independent realisations of the longitudinal profile generated with our model. The longitudinal profiles are rescaled in atmospheric depths by \xmax and the amplitude by \nmax. The colors show the different sampled values of $L$, spanning a wide range of $L$ values. }
    \label{fig:long_prof_priors}
\end{wrapfigure}

The model for the longitudinal profile also includes an additional subdominant component to capture longitudinal profiles that deviate from a standard Gaisser-Hillas-like profile. This component is modelled by a Correlated Field model~\cite{Arras:2022}, which encodes statistically homogeneous and isotropic correlations by a power spectrum. This power spectrum is modelled as a power-law with additional deviations modelled with an integrated Wiener process, allowing to learn the degree of correlation between neighbouring atmospheric bins (see e.g. \cite{niftyre} for a concrete example). This component is suppressed by a factor that is inferred with a prior of $\mathcal{N}(0, 10^{-3})$. We show different realisations of the longitudinal distribution in \Cref{fig:long_prof_priors}. \par 

\textbf{Radio Emission: } As in our previous work, we use SMIET \cite{Desmet:2025ufy} to describe the radio pulses (at the electric field level) produced from arbitrary longitudinal profiles. In SMIET, template pulses at each atmospheric slice $X$ are first produced by rescaling a ``sliced shower'', i.e., showers simulated with CoREAS \cite{Huege:2013vt} that measures the emission produced from each atmospheric slice, with ``spectral functions'' that capture the universal features of the emission. The target pulses are then synthesised by rescaling the template pulses with the given longitudinal profile. The performance of SMIET has already been validated in \cite{Desmet:2025ufy}. \par 

\begin{wrapfigure}[23]{L}{0.48\textwidth}
    \centering
    \includegraphics[width=\linewidth]{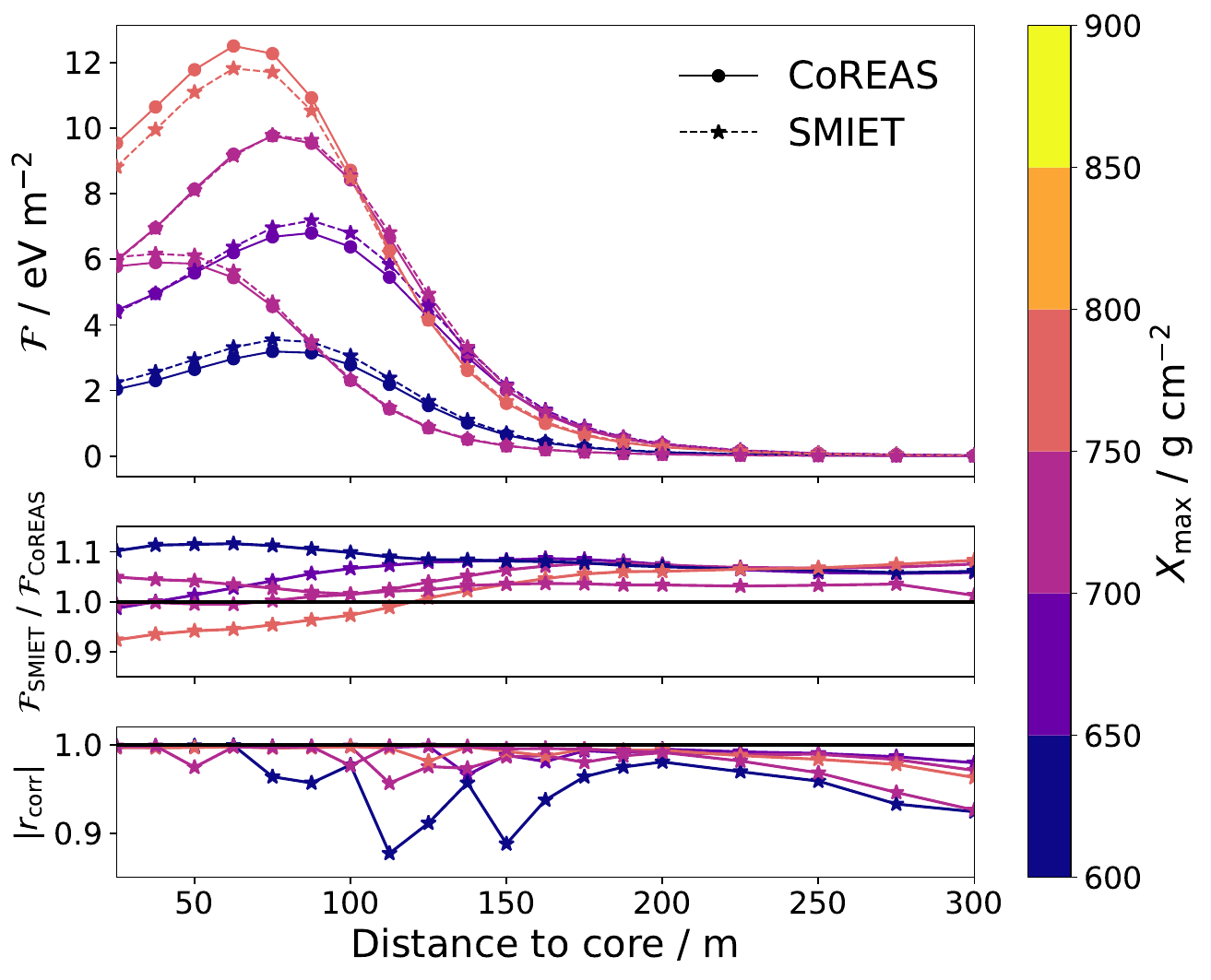}
    \caption{Top: energy fluence as a function of the distance to the shower core for CoREAS (solid, circle) and SMIET (dashed, star) simulated pulses, color-coded by the \xmax value. Middle: the fluence ratio between SMIET and CoREAS for each simulation. The amplitude of all simulated pulses with SMIET are within 6\% of that of CoREAS. Bottom: normalised Pearson's correlation coefficient, $|r_\mathrm{corr}|$ between the SMIET- and CoREAS-generated pulses. Most pulses have a correlation coefficient of > 0.9, indicating good agreement in the pulse shape.}
    \label{fig:coreas_smiet_validity}
\end{wrapfigure}

Previously, we have simulated a sliced shower specifically for each event given in the LOFAR simulation library \cite{Corstanje:2021kik}. This means that the event geometry, primary cosmic ray energy and local atmosphere model have been perfectly reproduced for each reconstruction. However, this method is not feasible in the far future, as the method is highly dependent on the reconstructed event properties, as well as the local atmospheric conditions. Furthermore, the accuracy of the synthesised pulse depends strongly on the \xmax from the sliced shower. \par 

To mitigate these issues, we have now produced a template library that can synthesise showers in a broad range of \xmax values and event geometries. To generate the library, we have simulated sliced showers in zenith bins of $[0^\circ, 60^\circ]$ in steps of $3^\circ$ for a single primary cosmic ray energy of $E_\mathrm{CR} = \SI{1e17}{\electronvolt}$, with protons as primary particles for all showers. This ansatz is valid as the templates are agnostic to both energy and mass composition, as this is taken into account through the amplitude and \xmax scaling of the profile, respectively. All simulations are performed with the U.S. standard atmosphere parametrised by Keilhauer with an observation level of $\SI{0}{\meter}$, and using the magnetic field from the LOFAR site. This configuration is chosen to replicate the one used to generate the spectral functions reported in \cite{Desmet:2025ufy}. We have also used Sibyll2.3d \cite{Riehn:2019jet} as the high-energy hadronic interaction model due to its fast computational speed. \par 

\xmax is a stochastic quantity, and as such it is not trivial to create templates with the desired coverage of $X_\mathrm{max}$. However, it is known that the atmospheric depth at the first point of interaction in the shower, $X_0$, which can be set manually for each simulation, linearly scales with $X_\mathrm{max}$. To generate this scaling relation, we follow the procedure similar to \cite{Bjarni:2021}. We first perform 100 CONEX simulations \cite{Bergmann:2006yz} at zenith angles of $0^\circ$, $15^\circ$, $30^\circ$, and $40^\circ$ for different values of the height of the first interaction, using the same configurations as above. After performing a linear fit between $X_0$ and $X_\mathrm{max}$, we use this relation to generate sliced showers, binned in \xmax from [600, 1200] $\si{\gram\per\centi\meter\squared}$ in steps of $\SI{100}{\gram\per\centi\meter\squared}$. While it is possible to generate this library on a star-shaped antenna layout, we opt only to synthesise on the \vvb axis for now to reduce computational complexity. \par 

To validate our template library, we compare our synthesised pulses with simulated pulses from CoREAS using the same configurations. The pulses are synthesised by taking the same longitudinal profile from each CoREAS simulation and applying SMIET with our template library, using the closest \xmax and zenith angle to the relevant simulation. \Cref{fig:coreas_smiet_validity} shows the resulting fluence distribution obtained from both CoREAS and SMIET. We also show the fluence ratio and the normalised Pearson's correlation coefficient between the pulses, which show agreement of around 6\% with each other, compatible to both the results from \cite{Desmet:2025ufy} and with intrinsic shower-to-shower fluctuations. \par 

In our model, the pulses are synthesised only on fixed antenna positions along the \vvb axis. This limits to 19 antennas used for the reconstruction. For simplicity, we also fix the arrival directions and core position of the shower, and as such we align the pulses with the arrival times from the simulated pulse, using the maximum position of the Hilbert envelope. In future works, we plan to extend this to realistic antenna layouts such as from LOFAR and SKA-Low, which can be incorporated with the Fourier-based pulse interpolation method \cite{Corstanje:2023vqp}. We also plan to include a timing-based model to additionally infer the arrival directions and core positions. \par 

\textbf{Antenna Response and Noise: } As with our previous works, we use the SKALA4.1 model \cite{Bolli2020} to model the response of all antennas at each antenna position. Through a convolution of the antenna response matrix with the electric field pulses from SMIET, we then have a description of the pulses at each antenna on the voltage level. The noise is modelled through a Gaussian distribution centered at zero with a diagonal covariance matrix with a noise root-mean-square value of $\sigma_V = \SI{2e-5}{\volt}$. This value is motivated from the simulated noise spectrum for the SKA-Low site, as used in \cite{Corstanje:2025wbc}. While it is possible to include sample-to-sample correlations of the noise \cite{Ravn:2025puy}, we do not yet include this in our model.

\textbf{Inference: } The inference is performed using NIFTy \cite{niftyre}, the numerical framework that utilises Information Field Theory. Unlike our previous work, which approximates the posterior as a Gaussian, we use the geoVI framework \cite{2021Entrp..23..853F}, which employs a coordinate transformation based on the local Fisher information metric to construct a more accurate variational approximation than a standard Gaussian. In this way, we directly capture the correlations between shower parameters. We use a total of 15 posterior samples and iterate up to 10 iterations, or until the inference has fully converged, which is set by ensuring that the residual between each time sample of the voltage trace of the reconstructed data from our model and the actual data, $\chi^2_\mathrm{res}$, normalised by $\mathrm{ndf}$, is $< 1.1$ for 3 consecutive iterations. The initial positions for \xmax and \nmax are set based on the energy-dependence of $\langle X_\mathrm{max} \rangle$ and $N_\mathrm{max}$, with $\sigma_{X_\mathrm{max}}^\mathrm{init} = \SI{50}{\gram\per\centi\meter\squared}$ and $\sigma_{\log E}^\mathrm{init} = 0.05$. For L and R, the initial positions are instead set such that we scatter the true value with $\sigma_L^\mathrm{init} = \SI{10}{\gram\per\centi\meter\squared}$ and $\sigma_R^\mathrm{init} = 0.05$.

\section{Validation on SMIET-generated dataset}

We apply our framework to realistic traces at the voltage level using simulated pulses from SMIET to not only benchmark the reconstruction performance but also check the self-consistency of our model. To use realistic longitudinal profiles, we take the profile generated for the LOFAR simulation library, consisting of 504 proton and 445 iron showers, with zenith angles from 0$^\circ$ - 40$^\circ$, azimuth angles from  0$^\circ$ - 360$^\circ$, and primary cosmic ray energies from $10^{16.5}$ - $10^{18.5}$ eV \cite{Corstanje:2021kik}. Each profile is used to synthesise pulses with SMIET, using the generated template library. The SKALA v4.1 antenna response function is applied to each electric field pulse to obtain the voltage-level signal. The data event is generated by adding Gaussian noise with $\sigma_V$ to each sample. To ensure good reconstruction quality, we take only events with $\chi^2_\mathrm{res} / \mathrm{ndf} < 1.05$ and with a signal-to-noise ratio of the strongest signal across all antennas $\mathrm{SNR}_\mathrm{peak} = \max_\mathrm{ant}(V) / \sigma_V > 1$. With this criterion, a total of 809 reconstructed events remain, which are used for analysis in this work.

\begin{figure}[!h]
    \centering
    \includegraphics[width=0.51\linewidth]{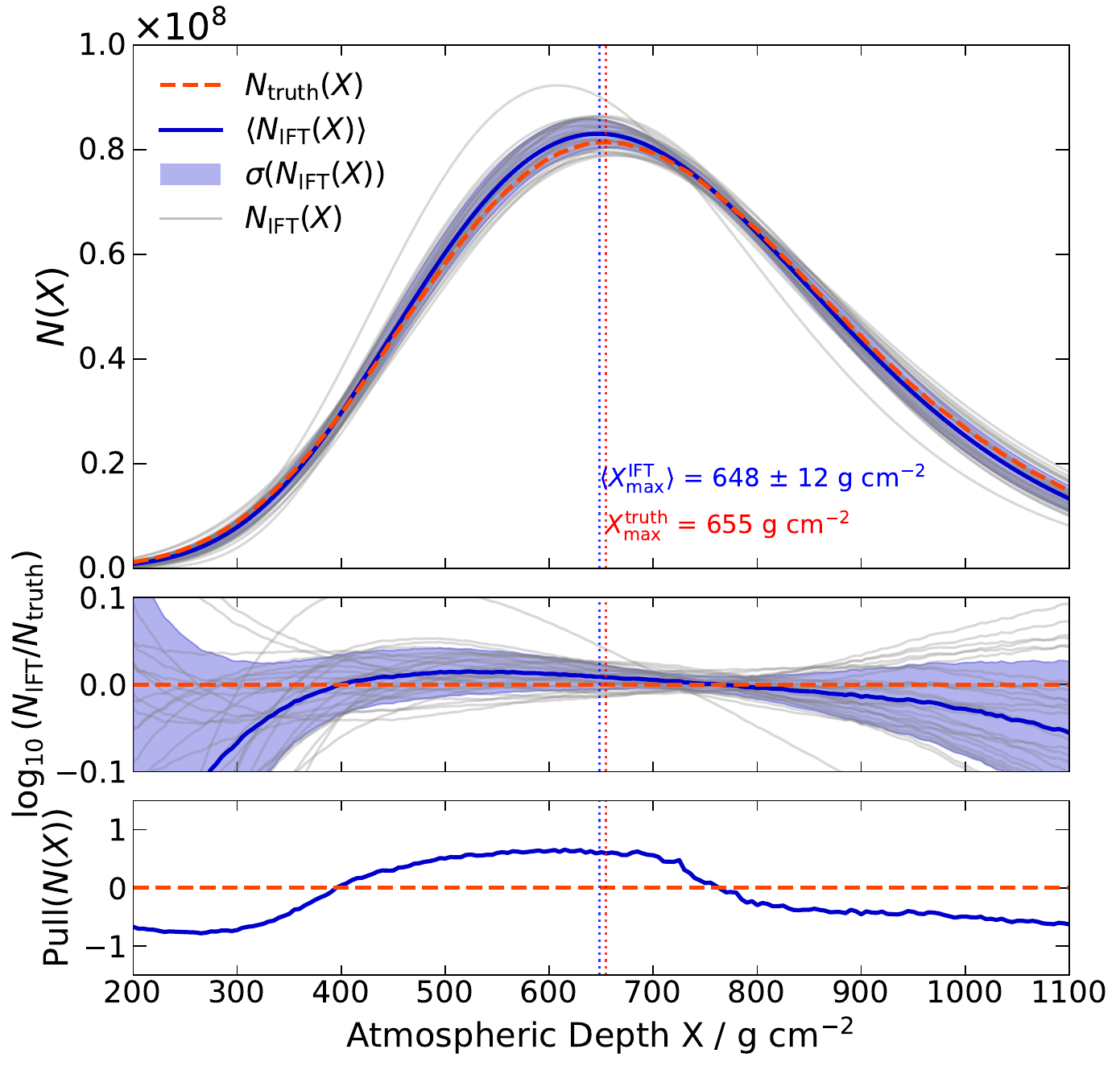}
    \includegraphics[width=0.48\linewidth]{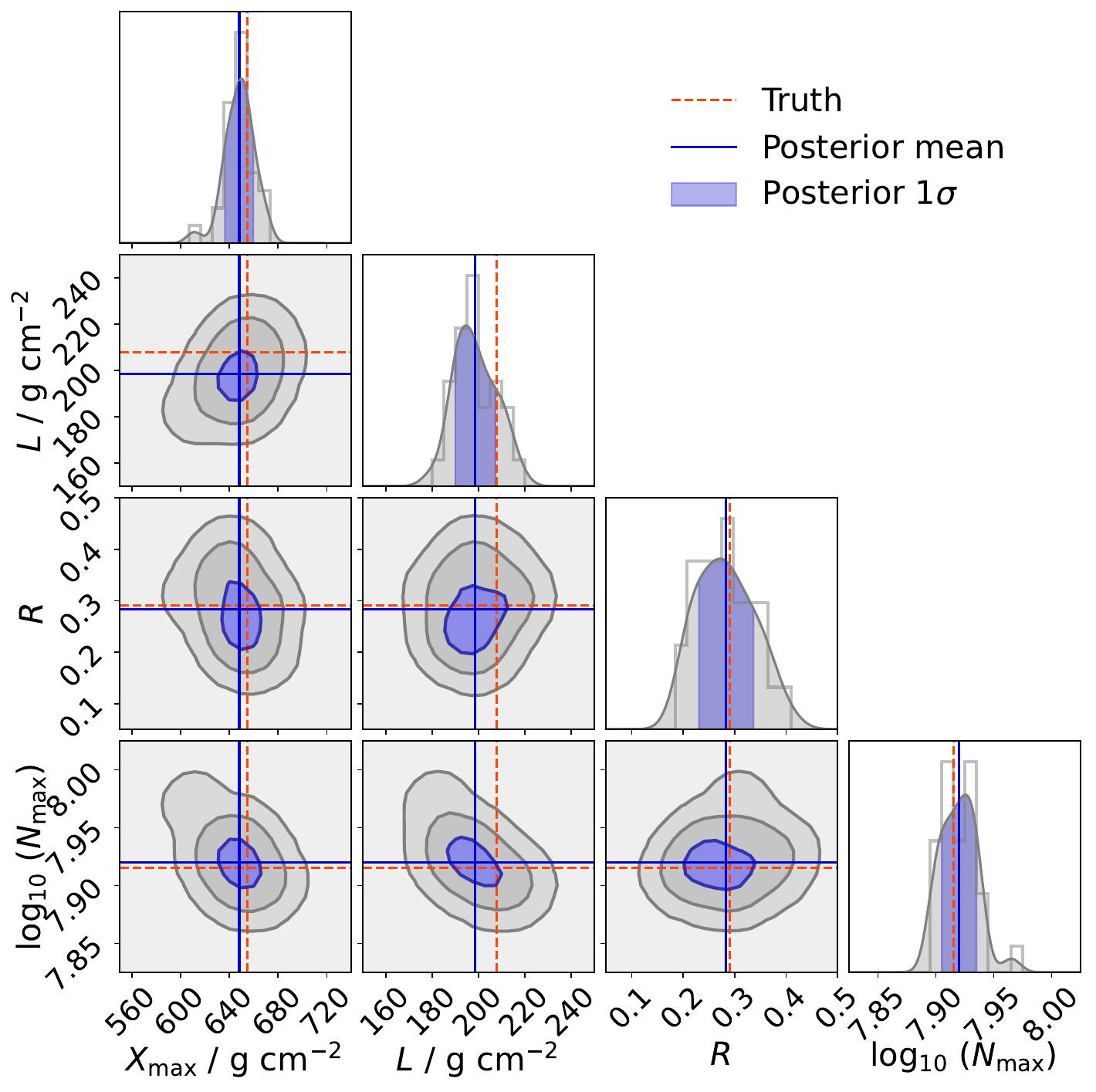}
    \caption{An example of a single event reconstruction using our framework. For all figures, we show the posterior mean (blue, solid), the standard deviation (blue, shaded), the individual posterior samples (gray), as well as the true value (orange, dashed). Left: The reconstructed longitudinal profile (top), the logarithmic ratio between the reconstructed and true profile (middle), and the pull (bottom). Right: Corner plot of the posterior samples of the shower parameters. The histograms are overlaid by the kernel density estimate (KDE) of the posterior distribution, and the 2-D contours show the smoothed 2-D KDE, where the contours are the 1, 2, and 3$\sigma$ values following a 2-D Gaussian distribution.}
    \label{fig:shower_reco}
\end{figure}

\begin{figure}[!h]
    \centering
    \includegraphics[width=0.43\linewidth]{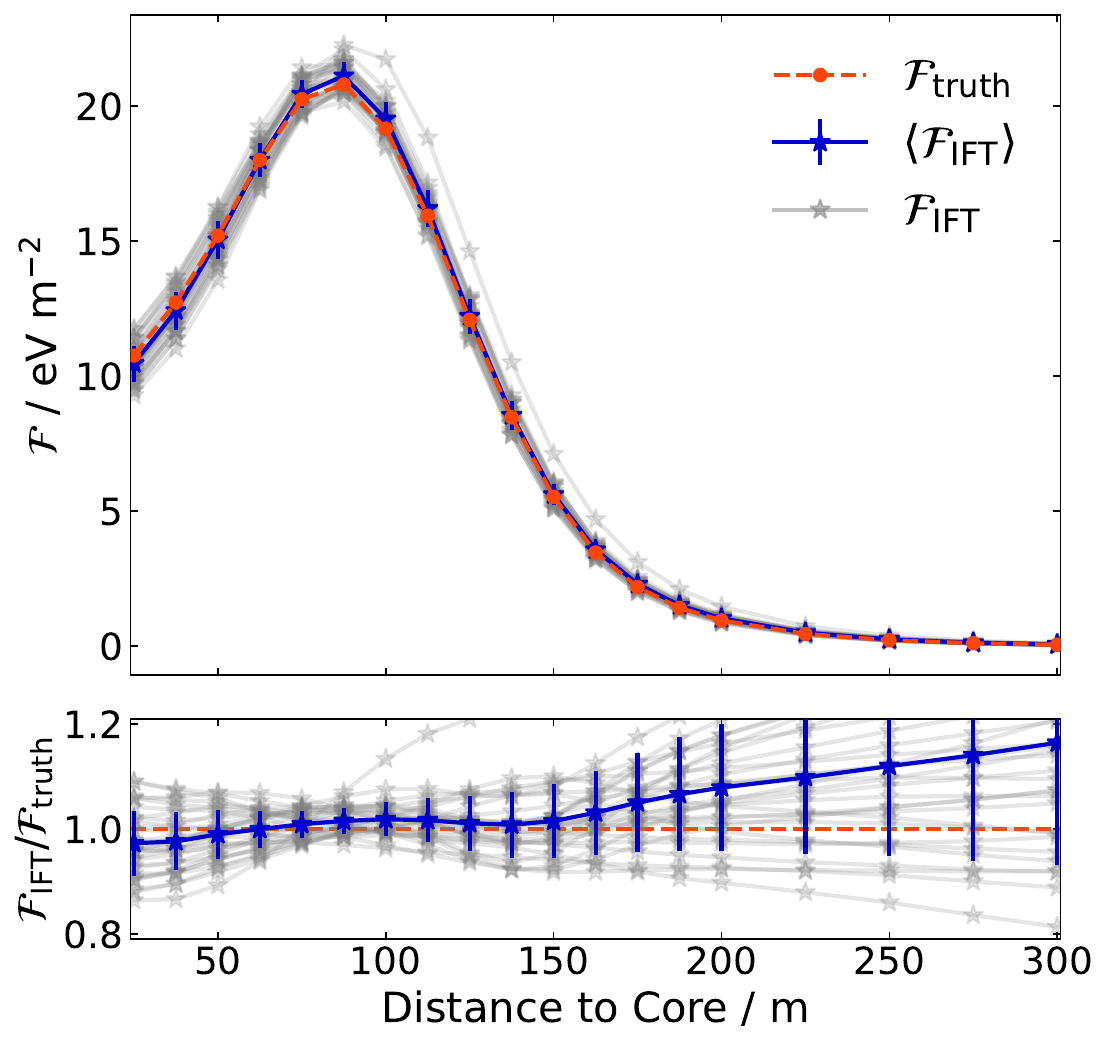}
    \includegraphics[width=0.56\linewidth]{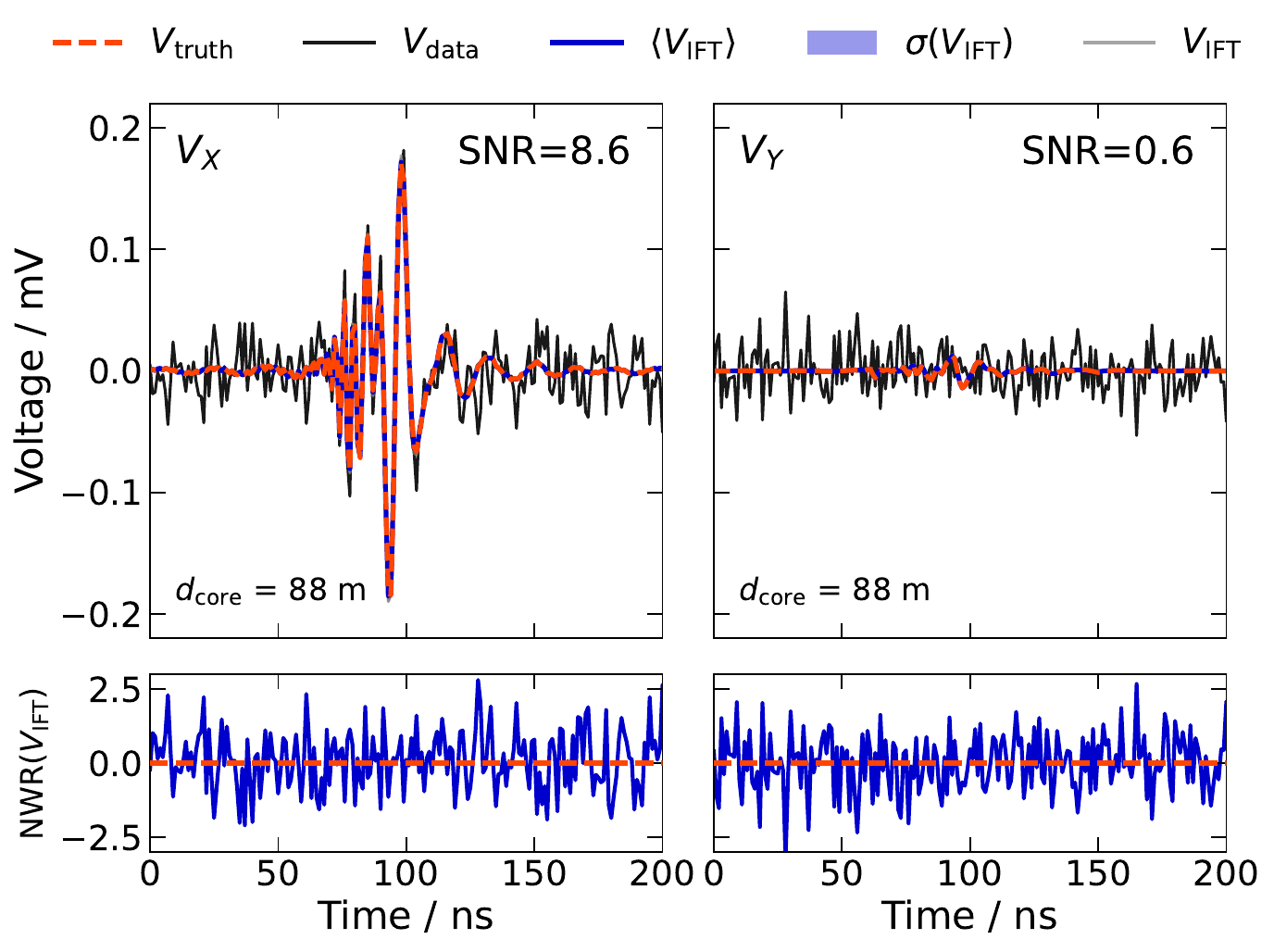}
    \caption{Same as \Cref{fig:shower_reco} but for signal-level quantities. Left: the energy fluence distribution as a function of antenna positions relative to the shower core (top) and the ratio between the reconstructed and true fluence (bottom). The reconstructed and true fluence values are marked by stars and circles, respectively. Right: The voltage traces in the $X$ and $Y$ polarization at distance of $d_\mathrm{core} = \SI{88}{\meter}$ from the core (top) and the noise-weighted residual (bottom). The mock data used in the reconstruction is shown in black.}
    \label{fig:fluence_efield_reco}
\end{figure}

\Cref{fig:shower_reco} shows an example reconstruction of the shower profile as well as the shower parameters for a single proton shower with $\theta = \SI{25}{\degree}$, $\phi = \SI{107}{\degree}$, $E_\mathrm{CR} = \SI{1.2e17}{\electronvolt}$, $X_\mathrm{max} = \SI{655}{\gram\per\centi\meter\squared}$ and $\mathrm{SNR}_\mathrm{peak} = 10.2$. We observe that here, most samples of the posterior distribution behave similarly, except at the earliest and latest stages of the shower evolution. Nevertheless, the shower is accurately reconstructed within 1$\sigma$ at all atmospheric depths, as shown by the pull distribution. We also observe a single outlier in the posterior samples; however, this individual sample does not contribute to the overall distribution, as the average behaviour of all samples is consistent with the reconstructed profile. The posterior distribution for the individual parameters shows that, not only can we reconstruct the true values within 1$\sigma$, but also we are able to capture the correlations between parameters without prior information of them. Notably, \nmax is strongly correlated with \xmax and $L$ due to the energy-dependence of these parameters.   \par 
\begin{figure}
    \centering
    \includegraphics[width=0.495\linewidth]{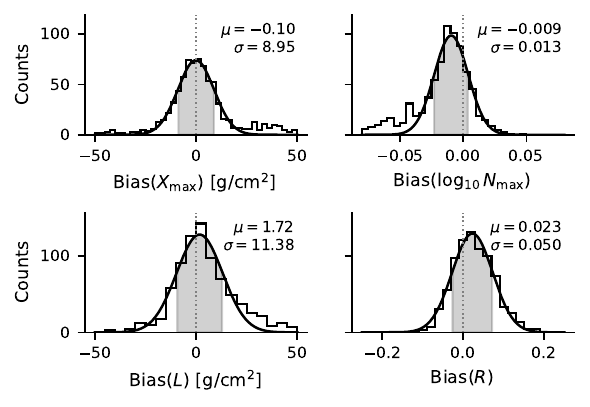}
    \includegraphics[width=0.495\linewidth]{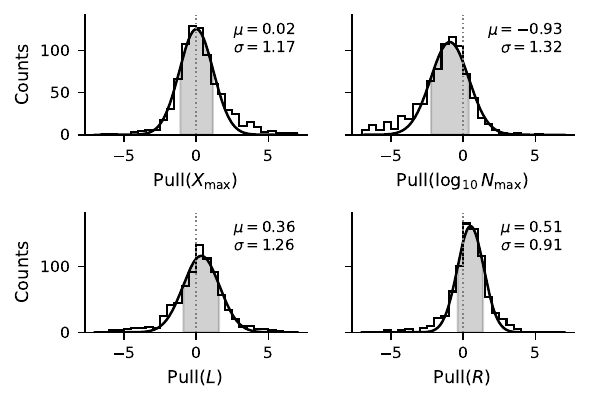}
    \caption{The performance of our reconstruction framework with the event set used in this work. Left: The bias distributions of $X_\mathrm{max}$, \lognmax, $L$, and $R$. The mean and resolution from fitting a standard normal distribution is shown for each parameter. The 1$\sigma$ band is shown in gray. Right: Same as the left figure but showing the pull distribution instead. }
    \vspace{-9mm}
    \label{fig:reco_performance}
\end{figure}

\Cref{fig:fluence_efield_reco} shows the signal-level quantities for the same event. The energy fluence distribution shows that only strong deviations originate from antennas far from the core, which are dominated by noise. Nevertheless, the fluence values are within 1$\sigma$ at all antenna positions. The voltage traces show that, as we reconstruct signals at both $X$ and $Y$ polarizations at all antennas simultaneously, we are able to reconstruct signals with a low SNR. The noise-weighted residual $\mathrm{NWR} = (V_\mathrm{IFT} - V_\mathrm{data}) / \sigma_V$ is consistently below 2.5, indicating that our model also accurately recovers the pulses from the synthesised data. \par 

\begin{wrapfigure}[14]{L}{0.48\textwidth}
    \centering
    \includegraphics[width=0.48\textwidth]{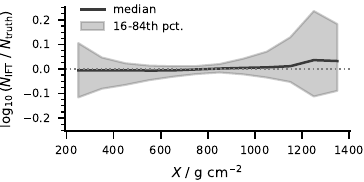}
    \caption{The median log ratio between reconstructed longitudinal profiles and the true profile for all reconstructed events (black, solid), for atmospheric depths in bins of $\SI{100}{\gram\per\centi\meter\squared}$. The 16th and 84th percentile is shown in gray. }
    \vspace{1mm}
    \label{fig:longprofile_rel_ratio}
\end{wrapfigure}

\Cref{fig:reco_performance} shows the bias and pull distributions for all shower parameters with all reconstructed events. We observe that all shower parameters are reconstructed with minimal bias, except $\log_{10} N_\mathrm{max}$, which shows a bias of 2\% in $N_\mathrm{max}$. The pull distributions show that, for all parameters, the uncertainties are well described in our framework. In particular, we achieve a resolution of $\sigma_{X_\mathrm{max}} = \SI{8.95}{\gram\per\centi\meter\squared}$, which is comparable to the fluence-based method shown in \cite{Corstanje:2025wbc} with a factor $\sim 2000$ fewer antennas. The distributions of $L$ and $R$ also show that, while they are more biased than $X_\mathrm{max}$, their uncertainties are well described. The resolutions are $\sigma_L < \SI{11.4}{\gram\per\centi\meter\squared}$ and $\sigma_R < 0.05$, respectively. We note that the tails of the \xmax and \lognmax are non-Gaussian, as the reconstruction behaves poorly for shallower showers with \xmax $< \SI{500}{\gram\per\centi\meter\squared}$. \par 

\Cref{fig:longprofile_rel_ratio} shows the median log ratio of the reconstruction longitudinal profile against the true profile for all events at atmospheric depth, binned by $\SI{100}{\gram\per\centi\meter\squared}$. The bias is consistently within $\sim 4\%$, until $X \sim \SI{1200}{\gram\per\centi\meter\squared}$, where it increases up to 20\%. However, the scatter is much larger at the tails of the profiles, so we do not expect to achieve good reconstruction performance there. Crucially, we are able to reconstruct the profile where the majority of the emission occurs with minimal bias.

\section{Conclusion}

In this work, we have shown a reconstruction framework to infer the longitudinal profile of cosmic ray air showers with realistic radio measurements, using 19 antennas aligned on the \vvb axis. We benchmarked the intrinsic reconstruction performance using a SMIET-generated dataset with $\sim 900$ events, using a CORSIKA-simulated profile. With these results, we show that we can reconstruct the entire longitudinal profile within uncertainty at all atmospheric depths, and capture non-trivial correlations between shower parameters. Our framework also recovers the shower parameters with minimal bias and good estimation of errors, yielding resolution of $< \SI{9}{\gram\per\centi\meter\squared}$, $< \SI{11.4}{\gram\per\centi\meter\squared}$ and $< 0.05$ for $X_\mathrm{max}$, $L$ and $R$, respectively. This shows that intrinsically, the framework achieves comparable reconstruction performance with other methods, even with only 19 antennas on the \vvb arm. The profile is also recovered with minimal bias of $< 4\%$ up to $X \sim \SI{1200}{\gram\per\centi\meter\squared}$. \par 

Nevertheless, we have only validated the self-consistency of our model, i.e., between SMIET and SMIET. As such, we will apply our framework to datasets generated from CoREAS simulations with measured noise using the NuRadioReco framework \cite{Glaser:2019rxw}. This can degrade the performance due to potential deviations between SMIET and CoREAS, but will confirm the validity of our framework on realistic air-shower measurements. Ultimately, we aim to apply our framework to a realistic antenna layout such as from LOFAR or SKA-Low, and include a timing-based model to additionally infer the arrival direction and core position of the shower. 

\let\oldbibliography\thebibliography
\renewcommand{\thebibliography}[1]{%
  \oldbibliography{#1}%
  \setlength{\itemsep}{1pt}%
}

{\footnotesize
\bibliographystyle{JHEP}

\providecommand{\href}[2]{#2}\begingroup\raggedright\endgroup

}
\newpage
\newcommand{\affilASTRON}{Netherlands Institute for Radio Astronomy (ASTRON), Dwingeloo, The Netherlands}
\newcommand{\affilCanTho}{Physics Education Department, School of Education, Can Tho University, Campus~II, 3/2 Street, Ninh Kieu District, Can Tho City, Viet Nam}
\newcommand{\affilCurtin}{International Centre for Radio Astronomy Research, Curtin University, Bentley, 6102, WA, Australia}
\newcommand{\affilDESY}{Deutsches Elektronen-Synchrotron DESY, Platanenallee~6, 15738 Zeuthen, Germany}
\newcommand{\affilErlangen}{Erlangen Centre for Astroparticle Physics, Friedrich-Alexander-Universit\"at Erlangen-N\"urnberg, 91058 Erlangen, Germany}
\newcommand{\affilGorlitz}{Deutsches Zentrum f\"ur Astrophysik, Postplatz~1, 02826 Görlitz, Germany}
\newcommand{\affilGroningen}{Kapteyn Astronomical Institute, University of Groningen, P.O.~Box 72, 9700 AB Groningen, Netherlands}
\newcommand{\affilHefei}{School of Astronomy and Space Science, University of Science and Technology of China, Hefei 230026, China}
\newcommand{\affilKanpur}{Department of Physics, Indian Institute of Technology Kanpur, Kanpur, UP-208016, India}
\newcommand{\affilKeyNanjing}{Key Laboratory of Modern Astronomy and Astrophysics, Nanjing University, Ministry of Education, Nanjing 210023, China}
\newcommand{\affilKIT}{Institut f\"ur Astroteilchenphysik, Karlsruhe Institute of Technology (KIT), P.O.~Box 3640, 76021 Karlsruhe, Germany}
\newcommand{\affilKhalifa}{Department of Physics, Khalifa University, P.O.~Box 127788, Abu Dhabi, United Arab Emirates}
\newcommand{\affilManchester}{Jodrell Bank Centre for Astrophysics, Department of Physics and Astronomy, University of Manchester, Manchester M13 9PL, UK}
\newcommand{\affilMaxPlanck}{Max-Planck Institut f\"ur Astrophysik, Karl-Schwarzschild-Str.~1, 85748 Garching, Germany}
\newcommand{\affilMunich}{Ludwig-Maximilians-Universit\"at M\"unchen (LMU), Geschwister-Scholl-Platz~1, 80539 M\"unchen, Germany}
\newcommand{\affilNanjing}{School of Astronomy and Space Science, Nanjing University, Nanjing 210023, China}
\newcommand{\affilNijmegen}{Department of Astrophysics/IMAPP, Radboud University Nijmegen, P.O.~Box 9010, 6500 GL Nijmegen, The Netherlands}
\newcommand{\affilNikhef}{Nikhef, Science Park Amsterdam, 1098 XG Amsterdam, The Netherlands}
\newcommand{\affilPurpleMt}{Key Laboratory of Dark Matter and Space Astronomy, Purple Mountain Observatory, Chinese Academy of Sciences, No.~10 Yuanhua Road, Nanjing, China}
\newcommand{\affilULB}{Universit\'e Libre de Bruxelles, Science Faculty CP230, B-1050 Brussels, Belgium}
\newcommand{\affilVUB}{Vrije Universiteit Brussel, Astrophysical Institute, Pleinlaan~2, 1050 Brussels, Belgium}
\newcommand{\affilXidian}{School of Electronic Engineering, Xidian University, No.2 South Taibai Road, Xi'an, China}
\newcommand{\affilSKA}{SKA Observatory, Jodrell Bank, Lower Withington, Macclesfield, SK11 9FT, UK}

\section*{Affiliations}
\scriptsize
\noindent
$^a$ {\affilKIT} \\
$^b$ {\affilErlangen} \\
$^c$ {\affilManchester} \\
$^d$ {\affilVUB} \\
$^e$ {\affilNijmegen} \\
$^f$ {\affilCurtin} \\
$^g$ {\affilMaxPlanck} \\
$^h$ {\affilMunich} \\
$^i$ {\affilGorlitz} \\
$^j$ {\affilGroningen} \\
$^k$ {\affilASTRON} \\
$^l$ {\affilPurpleMt} \\
$^m$ {\affilNikhef} \\
$^n$ {\affilDESY} \\
$^o$ {\affilKanpur} \\
$^p$ {\affilKhalifa} \\
$^q$ {\affilCanTho} \\
$^r$ {\affilSKA} \\
$^s$ {\affilNanjing} \\
$^t$ {\affilKeyNanjing} \\
$^u$ {\affilXidian} \\
$^v$ {\affilHefei} \\
$^w$ {\affilULB} \\

\section*{Acknowledgements} \noindent
SBo, AN and KT acknowledge funding through the Verbundforschung of the German Federal Ministry of Research, Technology and Space (BMFTR). PL, KW and MJ are supported by the Deutsche Forschungsgemeinschaft (DFG, German Research Foundation) – Projektnummer 531213488. MD is supported by the Flemish Foundation for Scientific Research (FWO-AL991). ST acknowledges funding from the Khalifa University RIG-S-2023-070 grant. SB acknowledges funding from the Medium-Scale Infrastructure program of the Flemish Foundation for Scientific Research (FWO). KM acknowledges funding from the Netherlands Research School for Astronomy (NOVA) Phase 6 Instrumentation Call.  The authors gratefully acknowledge the computing time provided on the high-performance computer HoreKa by the National High-Performance Computing Center at KIT (NHR@KIT). This center is jointly supported by the Federal Ministry of Education and Research and the Ministry of Science, Research and the Arts of Baden-W\"{u}rttemberg, as part of the National High-Performance Computing (NHR) joint funding program. HoreKa is partly funded by the German Research Foundation. VE and KT acknowledges funding through the German Federal Ministry of Education and Research for the project ErUM-IFT: Informationsfeldtheorie für Experimente an Großforschungsanlagen (Förderkennzeichen: 05D23EO1).

\end{document}